\def\Re{{\text{Re}}\,}
\def\kF{k_{\text{F}}}

\def\NF{N_{\text{F}}}

\def\epsilonF{\epsilon_{\text F}}

\def\be{\begin{equation}}
\def\ee{\end{equation}}
\def\bea{\begin{eqnarray}}
\def\eea{\end{eqnarray}}
\def\bse{\begin{subequations}}
\def\ese{\end{subequations}}

\def\Tmin{T_{\text{min}}}

\documentclass[prb,twocolumn,amsmath,amssymb,eqsecnum,preprintnumbers]{revtex4}
\usepackage{graphicx}
\usepackage{dcolumn}
\usepackage{bm}
\usepackage{bbm}
\usepackage{color}
\begin{document}
\title{Exact solution of the Boltzmann equation for low-temperature transport coefficients in metals II: Scattering by ferromagnons}
\author{J. Amarel$^1$, D. Belitz$^{1,2}$, and T.R. Kirkpatrick$^3$}
\affiliation{$^{1}$ Department of Physics and Institute for Fundamental Science,
                    University of Oregon, Eugene, OR 97403, USA\\
                   $^{2}$ Materials Science Institute, University of Oregon, Eugene,
                    OR 97403, USA\\
                  $^{3}$ Institute for Physical Science and Technology,
                    University of Maryland, College Park,
                    MD 20742, USA
            }
\date{\today}

\begin{abstract}
In a previous paper (Paper I) we developed a technique for exactly solving the linearized Boltzmann equation for the electrical and thermal 
transport coefficients in metals in the low-temperature limit. Here we adapt this technique to determine the magnon contribution to the 
electrical and thermal conductivities, and to the thermopower, in metallic ferromagnets. For the electrical resistivity $\rho$ at 
asymptotically low temperatures we find $\rho \propto \exp{(-\Tmin/T)}$, with $\Tmin$ an energy scale that results from the exchange 
gap and a temperature independent prefactor of the exponential. The corresponding result for the heat conductivity is 
$\sigma_h \propto T^3\,\exp{(\Tmin/T)}$, and thermopower is $S \propto T$. All of these results are exact, including the
prefactors. 
\end{abstract}


\maketitle

\section{Introduction}
\label{sec:I}

The scattering of conduction electron in metals by soft excitations, and the resulting temperature
dependence of the transport coefficients in the low-temperature ($T\to 0$) limit is an old problem.
The best known example is Bloch's $T^5$ law for the electrical resistivity due to the scattering by
acoustic phonons.\cite{Bloch_1930, Ziman_1960} In magnetic metals, the magnetic Goldstone
modes also contribute to the scattering. 
Magnons in antiferromagnets yield a $T^5$ contribution as phonons do,\cite{Yamada_Takada_1974, Ueda_1977}
whereas the corresponding result for helimagnets is $T^{5/2}$.\cite{Belitz_Kirkpatrick_Rosch_2006b}

Within magnetic systems, ferromagnets are a special case
in that only electron scattering between different sub-bands of the exchange-split conduction
band is possible. This leads to a lower limit $\Tmin$ on the energy transfer, with the temperature
scale $\Tmin$ determined by the exchange splitting and the spin-stiffness coefficient. For
temperatures large compared to $\Tmin$, Ueda and Moriya\cite{Ueda_Moriya_1975}
found that scattering by ferromagnons yield a $T^2$ contribution to the electrical resistivity.
For $T\ll \Tmin$ the electrical resistivity $\rho$
is exponentially small and has the form\cite{Bharadwaj_Belitz_Kirkpatrick_2014}
\bse
\label{eqs:1.1}
\be
\rho = \frac{m}{n e^2}\, T_1\,r(T)\,e^{-\Tmin/T}\ .
\label{eq:1.1a}
\ee
Here $m$, $n$, and $e$ are the electron mass, number density, and charge, respectively. $T_1$ is the magnetic
Debye temperature, and the dimensionless function $r$ is a power-law function of its argument.
By evaluating the Kubo formula in a conserving approximation, Ref.~\onlinecite{Bharadwaj_Belitz_Kirkpatrick_2014}
found $r(T) \propto (T/T_1)^2$.

All of the above results were obtained by solving either the linearized Boltzmann equation,
or an equivalent integral equation derived from the Kubo formula, in an uncontrolled approximation that 
replaces various energy-dependent relaxation rates by constants, see Ref.~\onlinecite{Mahan_2000} for
the electron-phonon case. Only very recently has it been shown that a mathematically rigorous solution of the
integral equation does indeed yield the Bloch $T^5$ law for the case of electron-phonon scattering.\cite{Amarel_Belitz_Kirkpatrick_2020}
In a previous paper\cite{Paper_I} (to be referred to as Paper I) we have simplified and extended the
method of Ref.~\onlinecite{Amarel_Belitz_Kirkpatrick_2020}. We have shown that the heat conductivity
and the thermopower can also be determined exactly, and we have applied the method to electron
scattering by antiferromagnons and helimagnons in addition to phonons. In all of these cases, it
turned out that the uncontrolled approximation affected the prefactor of the temperature dependence
of the transport coefficients, but the functional form of the various $T$-dependence was exact. It is
the purpose of the present paper to show that in ferromagnets the situation is different: 
an exact solution of the integral equation yields a prefactor $r$ in Eq.~(\ref{eq:1.1a}) that is
constant in the limit $T\to 0$,
\be
r(T) = 8 g_0\, \frac{T_0^3}{T_1\lambda} \left[1+ O(T\lambda/T_0^2)\right]\ . 
\label{eq:1.1b}
\ee
\ese
Here $T_0$ is a temperature scale that is closely related to $\Tmin$, $\lambda$ is the exchange splitting,
and $g_0$ is a dimensionless coupling constant that depends on the magnetization.
The corresponding result for the thermopower $S$ is
\be
S(T\to0) = \frac{-\pi^2}{6e}\,\frac{T}{\epsilonF}\ ,
\label{eq:1.2}
\ee
which is the same as for scattering by phonons, antiferromagnons, or helimagnons. The prefactors
in Eqs.~(\ref{eq:1.1b}) and (\ref{eq:1.2}) are exact. The result for the heat conductivity is
\be
\sigma_h(T\to 0) \propto (T^3/T_0^3)\,e^{\Tmin/T}\ ,
\label{eq:1.3}
\ee
but the numerical prefactor cannot be determined in closed form.

This paper is organized as follows. In Sec.~\ref{sec:II} we recall the linearized Boltzmann equations
for the magnon scattering contributions to the electrical and thermal resistivities, as well as for the thermopower. 
In Sec.~\ref{sec:III} we adapt the method from Paper I to solve the Boltzmann equation exactly in the
limit of asymptotically low temperature. We conclude in Sec.~\ref{sec:IV} with a summary and
a discussion of our results. Appendix~\ref{app:A} summarizes the derivation of the effective scattering
potential, and the derivation of the linearized Boltzmann equations from the Kubo formulas. Appendix~\ref{app:B}
lists various relaxation rates, and Appendix~\ref{app:C} contains some technical details regarding the
spectral analysis of the collision operator.

\section{Kinetic equations for transport coefficients in ferromagnets}
\label{sec:II}

\subsection{Energy scales, and transport coefficients}
\label{subsec:II.A}

We start by recalling some well known aspects of metallic ferromagnets. Their origins have been discussed
in detail in Ref.~\onlinecite{Bharadwaj_Belitz_Kirkpatrick_2014}, and we will just state the results.

In a ferromagnet, the magnetic order splits the conduction band into two sub-bands that are separated by the
exchange splitting $\lambda$ and can be indexed by the spin projection index $\sigma = \pm 1$. Let $\mu$ 
be the chemical potential, $\mu(T=0) = \epsilonF$ the Fermi energy, $\kF$ the Fermi wave number in the
non-magnetic state, and $m$ the electron effective mass. Then the Fermi wave numbers of the sub-bands are  
$\kF^{\sigma} = \kF \sqrt{1 + \sigma\lambda/\mu}$ and the corresponding densities of states are 
$\NF^{\sigma} = \kF^{\sigma} m/2\pi^2$. The Goldstone mode associated with the magnetic order is
the ferromagnon, with a frequency-momentum relation
\be
\omega_{\bm k} = D{\bm k}^2\ ,
\label{eq:2.1}
\ee
with $D$ the spin-wave stiffness coefficient.\cite{magnon_dispersion_footnote}
There are two relevant energy scales in addition to the Fermi energy. One is the magnetic Debye temperature
\be
T_1 = D\kF^2 \ .
\label{eq:2.2}
\ee
The other one is a temperature scale $T_0$ that is related to the minimum momentum transfer in
scattering processes between the two sub-bands that are mediated by ferromagnons,
\be
T_0 = \frac{1}{4}\,D (\kF^+ - \kF^-)^2 = \frac{1}{4}\,T_1 (\lambda/\epsilonF)^2
\label{eq:2.3}
\ee
A crucial feature of the coupling of ferromagnons to conduction electrons is that the magnons couple only
electrons in different sub-bands (`interband coupling'). This is in contrast to antiferromagnets and helimagnets, 
where the Goldstone modes can couple electrons in the same sub-band (`intraband coupling'), see Paper I. 
As a result, the energy scale $T_0$ plays an important role for transport processes: For temperatures
$T < T_0$ the magnon-induced scattering processes get frozen out, and all transport coefficients will 
show an exponential temperature dependence with the temperature scale set by $T_0$.

To define the transport coefficients we consider a mass current ${\bm J}$ and a heat current ${\bm J}_h$ driven by gradients of the electrochemical
potential $\bar\mu = \mu + eV$ and the temperature $T$, respectively. Here $e$ is the electron charge, and $V$ is the electric potential.
To linear order in the potential gradients the currents are determined by three independent transport coefficients 
(see, e.g., Ref.~\onlinecite{Mahan_2000}),
\bse
\label{eqs:2.4}
\bea
{\bm J} = -\frac{1}{T}\,L_{11} \bm\nabla\bar\mu - \frac{1}{T^2}\,L_{12}\,\bm\nabla T\ ,
\label{eq:2.4a}\\
{\bm J}_h = -\frac{1}{T}\,L_{12} \bm\nabla\bar\mu - \frac{1}{T^2}\,L_{22}\,\bm\nabla T\ .
\label{eq:2.4b}
\eea
\ese
An Onsager relation ensures that the same coefficient $L_{12}$ appears in both Eq.~(\ref{eq:2.4a})
and (\ref{eq:2.4b}). The electrical conductivity $\sigma$ is defined for the case of constant temperature
and constant chemical potential, via $e{\bm J} = -\sigma \bm\nabla V$. Analogously, the heat conductivity
$\sigma_h$ for a constant electrochemical potential is defined via ${\bm J}_h = -\sigma_h\bm\nabla T$.
We therefore have
\bse
\label{eqs:2.5}
\be
\sigma = \frac{e^2}{T}\,L_{11}\quad , \quad \sigma_h = \frac{1}{T^2}\,L_{22}\ .
\label{eq:2.5a}
\ee
The thermopower or Seebeck coefficient $S$ is defined in the absence of a mass current via $\bm\nabla V = S\,\bm\nabla T$,
and hence
\be
-eS = \frac{1}{T}\,\frac{L_{12}}{L_{11}}\ .
\label{eq:2.5c}
\ee
What is usually measured, rather than $\sigma_h$, is the heat conductivity $\kappa$ in the absence of a mass current. It is given by
\bea
\kappa &=& \frac{1}{T^2}\left(L_{22} - (L_{12})^2/L_{11}\right)
\nonumber\\
&=& \sigma_h - T\,S^2\,\sigma\ .
\label{eq:2.5d}
\eea
\ese
The three transport coefficients can all be expressed in terms of energy and spin dependent relaxation functions $\varphi_0$ and $\varphi_1$,
\bse
\label{eqs:2.6}
\bea
\sigma &=& \frac{e^2}{2m}\sum_{\sigma} n_{\sigma}\,\frac{1}{T} \int d\epsilon\,w(\epsilon)\,\varphi_0^{\sigma}(\epsilon)\ , \hskip 30pt
\label{eq:2.6a}\\
-ST\sigma/e &=& \frac{1}{2m}\sum_{\sigma} n_{\sigma}\,\frac{1}{T} \int d\epsilon\,w(\epsilon)\,\varphi_1^{\sigma}(\epsilon)\ ,
\label{eq:2.6b}\\
T\sigma_h &=& \frac{1}{2m}\sum_{\sigma} n_{\sigma}\,\frac{1}{T} \int d\epsilon\,w(\epsilon)\,\epsilon\,\varphi_1^{\sigma}(\epsilon)\ .
\label{eq:2.6c}
\eea
\ese
Here $n_{\sigma}$ is the electron density for spin projection $\sigma$. Here, and
throughout this paper, we denote by $\int d\epsilon$ a definite integral over all real values of $\epsilon$.
The weight function $w$ is given in terms of the Fermi function $f_0(x) = 1/(e^x+1)$, i.e., the equilibrium
distribution function of the electrons,
\bse
\label{eqs:2.7}
\be
w(\epsilon) = f_0(\epsilon/T)\left[1 - f_0(\epsilon/T)\right] = \frac{1}{\cosh^2(\epsilon/T)}
\label{eq:2.7a}
\ee
with a normalization
\be
\int d\epsilon\,w(\epsilon) = T\ .
\label{eq:2.7b}
\ee
\ese
$\varphi_0$ is dimensionally an inverse energy, and physically a relaxation time. $\varphi_1$ is dimensionless.
$\varphi_0$ and $\varphi_1$ are determined as the solutions of kinetic equations that we discuss next.

\subsection{Kinetic equations}
\label{subsec:II.B}

The integrals on the right-hand sides of Eqs.~(\ref{eqs:2.6}) can be written as Kubo expressions
for the particle-number current -- particle-number current, particle-number current -- heat current, and heat current -- heat current
correlations, respectively.\cite{Kubo_1957, Mahan_2000} The Kubo formulas give the exact linear
response of the system, and are very hard to evaluate. They are usually analyzed by means of a
conserving approximation that is equivalent to the linearized Boltzmann equation,\cite{Mahan_2000} and the non-equilibrium aspects of the bosons
(in our case, the ferromagnons) are ignored for simplicity.
Even this procedure leads to singular integral equations of Fredholm type that are hard to solve. The relevant
integral equation for the function $\varphi_0$ that determines the electrical conductivity was derived in 
Ref.~\onlinecite{Bharadwaj_Belitz_Kirkpatrick_2014}; the main steps of that derivation are summarized in
Appendix~\ref{app:A}. The analogous equations for $\varphi_1$ is obtained
via the same procedure by replacing the number current with the heat current. The result can be written in the form
\bse
\label{eqs:2.8}
\bea
\Lambda^{\sigma}(\epsilon)\,\varphi_0^{\sigma}(\epsilon) &=& -1\ ,
\label{eq:2.8a}\\
\Lambda^{\sigma}(\epsilon)\,\varphi_1^{\sigma}(\epsilon) &=& -\epsilon\ .
\label{eq:2.8b}
\eea
\ese
Here $\Lambda^{\sigma}(\epsilon)$ is a collision operator that is defined as
\be
\Lambda^{\sigma}(\epsilon) = \int du \sum_{\sigma'} \left[ K^{\sigma\sigma'}(\epsilon,u)\,R_{\epsilon\to u}^{\sigma\to\sigma'} - K_0^{\sigma\sigma'}(\epsilon,u) \right]\ ,
\label{eq:2.9}
\ee
with
\be
R_{\epsilon\to u}^{\sigma\to\sigma'} f^{\sigma}(\epsilon) = f^{\sigma'}(u)
\label{eq:2.10}
\ee 
for any spin-dependent function $f^{\sigma}(\epsilon)$.
The kernel $K$ has five contributions:\cite{Bharadwaj_Belitz_Kirkpatrick_2014, K34_footnote, magnon_dispersion_footnote}
\bea
K^{\sigma\sigma'}(\epsilon,u) &=& K_0^{\sigma\sigma'}(\epsilon,u) +  K_1^{\sigma\sigma'}(\epsilon,u)
  + K_2^{\sigma\sigma'}(\epsilon,u)
\nonumber\\
&&+  K_3^{\sigma\sigma'}(\epsilon,u) +  K_4^{\sigma\sigma'}(\epsilon,u)\ ,
\label{eq:2.11}
\eea   
from which one can construct relaxation rates
\be
\Gamma_n^{\sigma}(\epsilon) = \int du \sum_{\sigma'} K_n^{\sigma\sigma'}(\epsilon,u)\quad (n=0,1,2,3,4)\ ,
\label{eq:2.12} 
\ee

The basic ingredient of the kernel is $K_0$, all other parts can be expressed in terms of it. It can be 
written\cite{Bharadwaj_Belitz_Kirkpatrick_2014}
\bse
\label{eqs:2.13}
\be
K_0^{\sigma\sigma'}(\epsilon,u) = \NF^{\sigma'} \left[n_0\left(\frac{u-\epsilon}{T}\right) + f_0\left(\frac{u}{T}\right)\right] V_{\sigma\sigma'}(u-\epsilon)\ ,
\label{eq:2.13a}
\ee
where $n_0(x) = 1/(e^x-1)$ is the Bose distribution function. 
The effective potential $V$ reads
%
\bea
V_{\sigma\sigma'}(u) &=& -\sigma (1 - \delta_{\sigma\sigma'})\,\frac{g_0}{\NF}\,\Theta(-\sigma u)
\nonumber\\
&& \times \Theta(\vert u\vert - T_{\text{min}})\,\Theta(T_1 - \vert u\vert)\ .\qquad
\label{eq:2.13b}
\eea
\ese
The spin structure of this expression shows explicitly that the potential couples only electrons in
different sub-bands of the split conduction band. The lower energy cutoff $\Tmin$ is a result of
this structure, and it will obviously be closely related to $T_0$ as defined in Eq.~(\ref{eq:2.3}).
For $T_1\approx\lambda \ll \epsilonF$, which is always true in metals,
one finds
\be
T_{\text{min}}  = T_0\left[1 + T_0 \left(\frac{1}{T_1} + \frac{1}{\lambda}\right)\right] 
+ O\left(\frac{T_0^3}{T_1}, \frac{T_0^3}{T_1\lambda}, \frac{T_0^3}{\lambda^2}\right)\ .
\label{eq:2.14}
\ee
Here we ignore a spin dependence of the lower frequency cutoff that becomes relevant only at 
unrealizably low temperatures. The final step function in Eq.~(\ref{eq:2.13b}) reflects the upper
energy cutoff provided by the magnetic Debye temperature $T_1$, and $g_0$ is a dimensionless
coupling constant that is proportional to the residue of the ferromagnon pole, which in turn is
related to the magnetization.\cite{g0_footnote} 

The kernels $K_1$ through $K_4$ are related to $K_0$ via 
\bse
\label{eqs:2.15}
\bea
K_1^{\sigma\sigma'}(\epsilon,u) &=& \frac{-2\sqrt{T_0/T_1}}{1-4T_0/T_1}\,\sigma\,K_0^{\sigma\sigma'}(\epsilon,u)
\label{eq:2.15a}\\
K_2^{\sigma\sigma'}(\epsilon,u) &=& \frac{-2}{1-4T_0/T_1}\left(\frac{\vert u - \epsilon\vert}{T_1} - \frac{2 T_0}{T_1}\right)\,
\nonumber\\
&& \hskip 80pt \times K_0^{\sigma\sigma'}(\epsilon,u)\ ,
\nonumber\\
\label{eq:2.15b}\\
K_3^{\sigma\sigma'}(\epsilon,u) &=& \frac{\sqrt{T_0/T_1}}{1-4T_0/T_1}\left(\frac{u - \epsilon}{\lambda} + \frac{4\vert u - \epsilon\vert}{T_1}\,\sigma\right)\,
\nonumber\\
&& \hskip 80pt \times K_0^{\sigma\sigma'}(\epsilon,u)\ ,
\nonumber\\
\label{eq:2.15c}\\
K_4^{\sigma\sigma'}(\epsilon,u) &=& \frac{-2T_0/T_1}{1-4T_0/T_1}\,\frac{u - \epsilon}{\lambda}\,\sigma\,K_0^{\sigma\sigma'}(\epsilon,u).\ 
\label{eq:2.15d}
\eea
\ese
They give rise to five separate parts of the collision operator defined by
\bse
\label{eqs:2.16}
\bea 
\Lambda_0^{\sigma}(\epsilon) &=& \int du \sum_{\sigma'} \left[ K_0^{\sigma\sigma'}(\epsilon,u)\,R_{\epsilon\to u}^{\sigma\to\sigma'} - K_0^{\sigma\sigma'}(\epsilon,u) \right]\ ,
\nonumber\\
\label{eq:2.16a}\\
\Lambda_i^{\sigma}(\epsilon) &=& \int du \sum_{\sigma'} K_i^{\sigma\sigma'}(\epsilon,u)\,R_{\epsilon\to u}^{\sigma\to\sigma'}\quad (i=1,2,3,4)\ ,
\nonumber\\
\label{eq:2.16b}
\eea
and the full collision operator is given by
\be
\Lambda = \Lambda_0 + \Lambda_1 + \Lambda_2 + \Lambda_3 + \Lambda_4\ .
\label{eq:2.16c}
\ee
\ese

We have written the integral equations (\ref{eqs:2.8}) in a form that is standard in kinetic theory.\cite{Dorfman_vanBeijeren_Kirkpatrick_2020}
Their derivation from the Kubo formulas, and their relation to the form usually used in many-body theory, is explained in Appendix \ref{app:A}.

\subsection{Properties of the collision operator}
\label{subsec:II.C}

In order to consider the symmetry properties of the kernels $K_i$ we define a
spin-dependent weight function
\bse
\label{eqs:2.17}
\be
w_{\sigma}(\epsilon) = w(\epsilon)\,w_{\sigma}
\label{eq:2.17a}
\ee
with $w(\epsilon)$ from Eq.~(\ref{eq:2.7a}) and
\be
w_{\sigma} = \frac{1}{\nu}\,\left(\kF^{\sigma}/\kF\right) = \frac{1}{\nu}\,(1 + \sigma\lambda/\epsilonF)^{1/2}
\label{eq:2.17b}
\ee
where
\be
\nu = \sum_{\sigma} (1 + \sigma\lambda/\epsilonF)^{1/2}
\label{eq:2.17c}
\ee
so that 
\be
\int d\epsilon \sum_{\sigma} w_{\sigma}(\epsilon) = T\ .
\label{eq:2.17d}
\ee
\ese
The kernels then obey
\bse
\label{eqs:2.18}
\bea
w_{\sigma}(\epsilon)\,K_{0,2,4}^{\sigma\sigma'}(\epsilon,u) &=& w_{\sigma'}(u)\,K_{0,2,4}^{\sigma'\sigma}(u,\epsilon)\ ,
\label{eq:2.18a}\\
w_{\sigma}(\epsilon)\,K_{1,3}^{\sigma\sigma'}(\epsilon,u) &=& -w_{\sigma'}(u)\,K_{1,3}^{\sigma'\sigma}(u,\epsilon)\ ,\quad
\label{eq:2.18b}
\eea
\ese
We further define a scalar product in the space of real-valued functions by
\be
\langle\psi\vert\varphi\rangle = \int d\epsilon \sum_{\sigma} w_{\sigma}(\epsilon)\,\psi^{\sigma}(\epsilon)\,\varphi^{\sigma}(\epsilon)\ ,
\label{eq:2.19}
\ee
Equations~(\ref{eqs:2.18}) then imply that the collision operators $\Lambda_{0,2,4}$ are self-adjoint with respect to this
scalar product, whereas $\Lambda_{1,3}$ are skew-adjoint. It is further useful to define averages with respect to the
weight function $w_{\sigma}$ by
\be
\langle\varphi\rangle_{w_{\sigma}} = \frac{1}{T} \int d\epsilon\,w_{\sigma}(\epsilon)\,\varphi_{\sigma}(\epsilon) = \langle 1\vert\varphi\rangle/\langle 1\vert 1\rangle\ .
\label{eq:2.20}
\ee

The integral equations (\ref{eqs:2.8}) can now be written
\bse
\label{eqs:2.21}
\bea
\Lambda \vert\varphi_0\rangle &=& -\vert 1\rangle\ ,
\label{eq:2.21a}\\
\Lambda \vert\varphi_1\rangle &=& -\vert \epsilon\rangle\ ,
\label{eq:2.21b}
\eea
\ese
where $\vert 1\rangle$ represents the constant function that is identically equal to one, and
$\vert\epsilon\rangle$ represents the linear function $f(\epsilon) = \epsilon$. In particular,
the normalization of the weight function, Eq.~(\ref{eq:2.17d}), now takes the form
\be
\langle 1\vert 1\rangle = T\ ,
\label{eq:2.22}
\ee
and the transport coefficients from Eqs.~(\ref{eqs:2.16}) can be written
\bse
\label{eqs:2.23}
\bea
\sigma &=& \frac{n e^2}{mT}\,\langle\varphi_0\vert 1\rangle\ ,
\label{eq:2.23a}\\
-S T\sigma/e &=& \frac{n}{mT}\,\langle\varphi_1\vert 1\rangle\ ,
\label{eq:2.23b}\\
T\sigma_h &=& \frac{n}{mT}\,\langle\varphi_1\vert\epsilon\rangle\ .
\label{eq:2.23c}
\eea
\ese

$\Lambda_0$ has a zero eigenvalue with the constant function as the eigenfunction. This is
true by construction: From Eq.~(\ref{eq:2.16a}) we immediately obtain
\be
\Lambda_0 \vert 1\rangle = 0\ .
\label{eq:2.24}
\ee
The physical meaning of this zero eigenvalue is the approximate conservation law for the
electron momentum in the limit $T\to 0$, where the momentum transfer due to magnons is
frozen out. The zero eigenvalue has multiplicity one, and all other eigenvalues are negative.
The proof of these statements is exactly analogous to the 
proof given in Sec. II.B.3 of Paper I for the electron-phonon case.

All of the above is an obvious generalization of the formalism developed in Paper I. Also following
Paper I, we assume that the collision operator $\Lambda$ has a spectral representation
\be
\Lambda = \sum_n \mu_n\,\frac{\vert e_n\rangle\langle e_n\vert}{\langle e_n\vert e_n\rangle}
\label{eq:2.25}
\ee
with eigenvalues $\mu_n$ and a complete orthogonal set of right eigenvectors $\vert e_n\rangle$
and left eigenvectors $\langle e_n\vert$,\cite{LR_footnote} so the unit operator is represented by
\be
\mathbbm{1} = \sum_n \frac{\vert e_n\rangle\langle e_n\vert}{\langle e_n\vert e_n\rangle}\ .
\label{eq:2.26}
\ee

In the following section we will use this spectral representation to construct exact solutions
of the integral equations (\ref{eqs:2.21}). As in Paper I, we will need to distinguish between
the `hydrodynamic' part of the function $\varphi_1$, which is related to the perturbed zero
eigenvalue of the collision operator, and the `non-hydrodynamic' or `kinetic' part that is unrelated
to the zero eigenvalue. To lowest order in our expansion, the kinetic part is given by $\vert h\rangle$,
which is the solution of
\be
\Lambda_0 \vert h\rangle = -\vert\epsilon\rangle\ .
\label{eq:2.27}
\ee
This equation has a solution since the inhomogeneity is orthogonal to the zero eigenvector, $\langle\epsilon\vert 1\rangle = 0$.

\section{Solutions of the kinetic equations}
\label{sec:III}

In this section we construct formally exact solutions of the kinetic equations (\ref{eqs:2.8}). Our technique
for doing so is modeled after the analysis of the electron-phonon scattering problem in Paper I, which in
turn is based on a mathematically rigorous treatment that was given in Ref.~\onlinecite{Amarel_Belitz_Kirkpatrick_2020}. 
What makes the exact solution possible is the zero eigenvalue of the collision operator $\Lambda_0$, see
Eq.~(\ref{eq:2.24}). The other parts of the collision operator in Eq.~(\ref{eq:2.16c}) perturb the zero eigenvalue.
If $T$ and $T_0$ are both small compared to the magnetic Debye temperature $T_1$, these perturbations are
small and allow for a controlled determination of the smallest eigenvalue, which dominates the transport coefficients.
A complication compared to Paper I arises from the fact that the potential that governs the effective
electron-electron interaction, $V$ in Eq.~(\ref{eq:2.13b}), is gapped, and care must be taken to distinguish
between powers of $T$ and powers of $\Tmin \approx T_0$.

\subsection{Solutions of the integral equations}
\label{subsec:III.A}

\subsubsection{Scaling considerations}
\label{subsubsec:III.A.1}

We are interested in the behavior in the low-temperature regime defined by $T \ll T_0$. Accordingly, we introduce a small parameter $\alpha$
that scales as $\alpha \sim \sqrt{T/T_1} \ll 1$ that counts powers of temperature. Only even powers of $\alpha$
will occur in the low-temperature expansion. In addition, we assume that $\lambda/\epsilonF \approx T_1/\epsilonF \ll 1$,
and associate another small counting parameter $\beta \sim T_1/\epsilonF \sim \sqrt{T_0/T_1}$ with this energy
ratio. (This is true in metals, but not necessarily in, e.g., magnetic semiconductors.) $\lambda$ and $T_1$ are physically 
different energy scales, but their values are usually of the same order and we will not distinguish between them for scaling 
purposes. An inspection of the kernels, Eqs.~(\ref{eqs:2.13} - \ref{eqs:2.15}), shows that the collision operators $\Lambda_n$
scale, to leading order, as $\Lambda_n \sim \beta^{n+2}$. Futhermore, matrix elements that involve the vector $\vert\epsilon\rangle$
scale as $\langle\epsilon\vert\Lambda_n\vert\epsilon\rangle/T \sim \beta^{n+2}$, since only the $\epsilon$-integration
measure scales as the temperature, which gets canceled by the normalization factor $1/T$. Corrections to the leading
scaling behavior involve powers of $\alpha \sim \sqrt{T/T_1}$, which for $T\ll T_0$ are small compared to 
$\beta \sim \sqrt{T_0/T_1}$ to the same power. As a simple example, consider the average of the rate
$\Gamma_0$, Eq.~(\ref{eq:2.12}), that is calculated in Appendix~\ref{app:A}. The result is
\bse
\label{eqs:3.1}
\be
\langle\Gamma_0\rangle_{w_{\sigma}} \propto (\Tmin + T) e^{-\Tmin/T}\ ,
\label{eq:3.1a}
\ee
and the leading scaling behavior thus is
\be
\langle\Gamma_0\rangle_{w_{\sigma}} \sim \beta^2 + \alpha^2\ .
\label{eq:3.1b}
\ee
\ese
In general, for temperatures $T\ll T_0$ the $T_0$-scaling will dominate, and the only temperature dependence
of observables, other than the leading exponential one, will result from factors such as $\langle\epsilon^2\rangle_{w{\sigma}} \propto T^2$
that do not involve the collision operator. There is, however, one exception to this conclusion: Suppose an observable
$\cal O$ scales as $\beta^{2n}$ plus corrections, but the leading term has a zero prefactor:
\be
{\cal O} \sim 0\times\beta^{2n} + \beta^{2n-2}\alpha^2 + \beta^{2n+2} + O(\beta^{2n-4}\alpha^4,\beta^{2n}\alpha^2)\ .
\label{eq:3.2}
\ee
 Then $\alpha^2$ competes with $\beta^4$ rather than $\beta^2$, and thus the leading $\alpha$-correction dominates
 over the leading nonzero $\beta$-scaling for temperatures $T_0^2/T_1 < T < T_0$, and has to be kept. As we will see,
 this does indeed happen. In all cases where the coefficient of the leading $\beta$-scaling term is nonzero, on the other
 hand, we can neglect all temperature corrections. 
 
We now write the collision operator as
\be
\Lambda = \beta^2 \Lambda_0 + \beta^3 \Lambda_1 + \beta^4 \Lambda_2 + \beta^5 \Lambda_3 + \beta^6 \Lambda_4\ ,
\label{eq:3.3}
\ee
where powers $\beta^n$ imply that the corresponding part of the collision operator scales {\em at least}
as $(T_1/\epsilonF)^n \sim (\lambda/\epsilonF)^n$, or $(T_0/T_1)^{n/2}$. Forming matrix elements with these
collision operators will lead to $\alpha$-corrections to the leading scaling behavior, which we will keep as
needed. These scaling behaviors all pertain to the prefactor of the exponential $\exp(-T_{\text{min}}/T)$.
In the end, we will put $\beta = 1$.

\subsubsection{The inverse collision operator}
\label{subsubsec:III.A.2}

We now proceed in analogy to Paper I. That is, we expand the lowest eigenvalue $\mu_0$ and the corresponding
eigenvector $\vert e_0\rangle$ in power series in $\beta$, 
\bse
\label{eqs:3.4}
\bea
\mu_0 = \beta^2 \mu_0^{(0)} + \beta^3\mu_0^{(1)} + \beta^4 \mu_0^{(2)} + O(\beta^5)\ ,
\label{eq:3.4a}\\
\vert e_0\rangle = \vert e_0^{(0)}\rangle + \beta  \vert e_0^{(1)}\rangle + \beta^2  \vert e_0^{(2)}\rangle + O(\beta^3)
\label{eq:3.4b}
\eea
\ese
and solve the eigenproblem 
\be
\Lambda \vert e_0\rangle = \mu_0 \vert e_0\rangle
\label{eq:3.5}
\ee
order by order in $\beta$. This will allow us to construct the leading behavior of the inverse collision operator,
$\Lambda^{-1}$, which in turn will yield the leading contributions to the solutions of the integral equations
(\ref{eqs:2.21}). As mentioned above, the various parts of the collision operator,
and hence the eigenvalues and eigenvectors, contain powers of $\sqrt{T_0/T_1}$ equal to or higher than the one
indicated by the power of $\beta$ in Eqs.~(\ref{eq:3.3}) and (\ref{eqs:3.4}). Rather than keeping explicit powers
of $\beta$ everywhere, we will therefore use $\beta$ interchangeably with
$\sqrt{T_0/T_1}$, and $\alpha$ with $\sqrt{T/T_1}$, mostly to indicate leading scaling behavior, and higher-order
corrections. At all times we will maintain a systematic double expansion in powers of $\beta \sim \sqrt{T_0/T_1}$
and $\alpha \sim \sqrt{T/T_1}$.

Identities that will be useful in this context are
\bse
\label{eqs:3.6}
\bea
\Lambda_0 \vert 1\rangle &=& 0\ ,
\label{eq:3.6a}\\
\Lambda_1\vert 1\rangle &=& \frac{\sqrt{T_0/T_1}}{1-4T_0/T_1}\,\Lambda_0 \vert \sigma \rangle
\label{eq:3.6b}\\
\Lambda_1\vert \sigma\rangle &=&  \frac{-\sqrt{T_0/T_1}}{1-4T_0/T_1}\,\sigma\Lambda_0 \vert \sigma \rangle
\label{eq:3.6c}\\
\Lambda_2\vert 1\rangle &=& \frac{2/T_1}{1-4T_0/T_1}\,\sigma\Lambda_0\vert\epsilon\rangle - \frac{2T_0/T_1}{1-4T_0/T_1}\,\sigma\Lambda_0\vert\sigma\rangle
\nonumber\\
\label{eq:3.6d}\\
\Lambda_2\vert\sigma\rangle &=&  \frac{-2/T_1}{1-4T_0/T_1}\,\Lambda_0\vert\epsilon\rangle + \frac{2T_0/T_1}{1-4T_0/T_1}\,\Lambda_0\vert\sigma\rangle
\nonumber\\
\label{eq:3.6e}\\
\Lambda_3\vert 1\rangle  &=& \frac{\sqrt{T_0/T_1}}{1-4T_0/T_1}\left(\frac{1}{\lambda} - \frac{4}{T_1}\right)\Lambda_0\vert\epsilon\rangle\ .
\label{eq:3.6f}
\eea
\ese
Here $\vert\sigma\rangle$ represents the function $f^{\sigma}(\epsilon) = \sigma$, and $\sigma\Lambda_0$
denotes the collision operator constructed from the kernel $\sigma K_0^{\sigma\sigma'}(\epsilon,u)$. 

To lowest (i.e., quadratic) order in $\beta$ we have, by construction,
\bse
\label{eqs:3.7}
\bea
\mu_0^{(0)} &=& 0\ ,
\label{eq:3.7a}\\
\vert e_0^{(0)}\rangle &=& \vert 1\rangle\quad,\quad \langle e_0^{(0)}\vert = \langle 1\vert\ .
\label{eq:3.7b}
\eea
\ese
This is the zero eigenvalue that was mentioned in Sec.~\ref{subsec:II.C}.

To next-leading order the eigenequation reads
\be
\Lambda_0 \vert e_0^{(1)}\rangle + \Lambda_1 \vert 1\rangle = \mu_0^{(1)} \vert 1\rangle\ .
\label{eq:3.8}
\ee
Multiplying from the left with $\langle 1\vert$ yields
\bse
\label{eqs:3.9}
\be
\mu_0^{(1)}= 0\ .
\label{eq:3.9a}
\ee
To find the corresponding eigenvector we use Eq.~(\ref{eq:3.6b}), which implies
\bea
\vert e_0^{(1)}\rangle &=& \frac{-\sqrt{T_0/T_1}}{1-4T_0/T_1}\,\vert\sigma\rangle + \frac{1}{2}\,\frac{1-\sqrt{1-4T_0/T_1}}{1-4T_0/T_1}\,\vert 1\rangle\ .
\nonumber\\
&=& -\left(\frac{T_0}{T_1}\right)^{1/2}\left[1 + O(\beta^2\right)] \vert\sigma\rangle + \frac{T_0}{T_1}\left[1 + O(\beta^2)\right] \vert 1\rangle\ .
\nonumber\\
\label{eq:3.9b}
\eea
\ese
Note that an arbitrary
multiple of the zero eigenfunction $\vert 1\rangle$ can be added to $\vert e_0^{(1)}\rangle$; Eq.~(\ref{eq:3.9b})
reflects the fact that  $\vert e_0^{(1)}\rangle$ should be orthogonal to $\vert 1\rangle$. We use the same notation
for the right and left eigenfunctions as in Sec.~\ref{subsec:II.C}, and it comes with the same caveats.\cite{LR_footnote} 
Accordingly, the left even eigenvectors $\langle e_0^{(2n)}\vert$ represent the same functions as the corresponding
right eigenvectors $\vert e_0^{(2n)}\rangle$, whereas the left odd eigenvectors $\langle e_0^{(2n+1)}\vert$ represent
minus the functions represented by the  $\vert e_0^{(2n+1)}\rangle$. This is a consequence of the skew-adjointness
of the operators $\Lambda_{2n+1}$. 

To order $\beta^4$, the equation
\be
\Lambda_0 \vert e_0^{(2)}\rangle + \Lambda_1 \vert e_0^{(1)}\rangle + \Lambda_2 \vert 1\rangle = \mu_0^{(2)} \vert 1\rangle
\label{eq:3.10}
\ee
yields, for the eigenvalue,
\bse
\label{eqs:3.11}
\be
\mu_0^{(2)} = \langle\Gamma_2\rangle_{w_{\sigma}} - \frac{2 T_0/T_1}{(1-4 T_0/T_1)^2} \langle \Gamma_0\rangle_{w_{\sigma}}\ .
\label{eq:3.11a}
\ee
A calculation of the average rates, see Appendix~\ref{app:A}, shows that the leading contributions, i.e., the ones that scale as 
$\beta^4 \sim (T_0/T_1)^2$, cancel between the two terms. This is also obvious from the relation (\ref{eq:C.3a}) 
(see also Eq.~(\ref{eq:2.15b})) between the kernels $K_2$ and $K_0$, since the factor $\vert u-\epsilon\vert$ in $K_2$ turns 
into $T_0/T_1$ to leading order. However, the leading corrections, which scale as $\beta^2\alpha^2 \sim T_0 T/T_1^2$, do not cancel, 
and we have
\be
\mu_0^{(2)} = 8 g_0\,T_0 \left[\frac{T}{T_1} + O(\beta^4)\right] e^{-T_{\text{min}}/T}\ .
\label{eq:3.11b}
\ee
The full eigenvector $\vert e_0^{(2)}\rangle$ is not needed, but we do need its overlap with $\vert\epsilon\rangle$, which
vanishes, see Appendix~\ref{app:C},
\be
\langle\epsilon\vert e_0^{(2)}\rangle = 0\ .
\label{eq:3.11c}
\ee
\ese

The cancellation of the leading terms in $\mu_0^{(2)}$ gives rise to the competition mechanism explain in conjunction with Eq.~(\ref{eq:3.2})
and forces us to go to higher order. At $O(\beta^5)$ we have
\be
\Lambda_0 \vert e_0^{(3)}\rangle + \Lambda_1 \vert e_0^{(2)}\rangle + \Lambda_2 \vert e_0^{(1)}\rangle + \Lambda_3 \vert 1\rangle
   =
 \mu_0^{(2)} \vert e_0^{(1)}\rangle + \mu_0^{(3)} \vert 1\rangle\ . 
\label{eq:3.12}
\ee
Multiplying from the left with $\langle 1\vert$, and using Eqs.~(\ref{eq:3.8}) and (\ref{eq:3.10}), we can eliminate the matrix
element that involves the unknown eigenvector $\vert e_0^{(2)}\rangle$. Using the skew-adjointness of $\Lambda_1$ and $\Lambda_3$ 
we then find
\bse
\label{eqs:3.13}
\be
\mu_0^{(3)} = 0\ .
\label{eq:3.13a}
\ee
For the corresponding eigenvector we will again need only its overlap with $\vert\epsilon\rangle$. We find, see Appendix~\ref{app:C},
\be
\langle\epsilon\vert e_0^{(3)}\rangle = -b \langle\epsilon\vert\epsilon\rangle + O(T^4)\ ,
\label{eq:3.13b}
\ee
where
\be
b =  \frac{1}{\lambda}\,\left(\frac{T_0}{T_1}\right)^{1/2} + O(\beta^2) = \frac{1}{2\epsilonF} + O(\beta^2)\ .
\label{eq:3.13c}
\ee
For the purpose of calculating the overlap $\langle\epsilon\vert e_0\rangle$, the right and left eigenvectors at cubic order are thus
adequately represented by
\bea
\vert e_0^{(3)}\rangle &\approx& -b \vert\epsilon\rangle\ ,
\label{eq:3.13d}\\
\langle e_0^{(3)}\vert &\approx& b \langle\epsilon\vert\ .
\label{eq:3.13e}
\eea
\ese

Finally, at order $\beta^6$ the integral equation
\begin{widetext}
\be
\Lambda_0 \vert e_0^{(4)}\rangle + \Lambda_1 \vert e_0^{(3)}\rangle + \Lambda_2 \vert e_0^{(2)}\rangle + \Lambda_3 \vert e_0^{(1)}\rangle + \Lambda_4 \vert 1\rangle =
 \mu_0^{(2)} \vert e_0^{(2)}\rangle + \mu_0^{(4)} \vert 1 \rangle
\label{eq:3.14}
\ee
yields
\bea
\mu_0^{(4)} \langle 1\vert 1\rangle &=& \langle 1\vert \Lambda_1\vert e_0^{(3)}\rangle + \langle 1\vert\Lambda_2\vert e_0^{(2)}\rangle
   + \langle 1\vert\Lambda_3\vert e_0^{(1)}\rangle + \langle 1\vert\Lambda_4\vert 1\rangle
\nonumber\\
&=&
\langle e_0^{(2)}\vert\Lambda_1\vert e_0^{(1)}\rangle + \langle e_0^{(2)}\vert\Lambda_2\vert 1\rangle 
   + \langle e_0^{(1)}\vert\Lambda_2\vert e_0^{(1)}\rangle
+ 2 \langle e_0^{(1)}\vert \Lambda_3\vert 1\rangle + \langle 1\vert\Lambda_4\vert 1\rangle\ .
\label{eq:3.15}
\eea
\end{widetext}
Here we have used Eq.~(\ref{eq:3.12}) to write all matrix elements in terms
of the eigenvector up to second order only. We again observe that, upon doing the integrals, and to leading order in our expansion
in powers of $\beta$, the term $\vert u - \epsilon\vert/T_1$ in the definition
of $K_2$, Eq.~(\ref{eq:2.15b}), turns into $T_0/T_1$. As a result, the sum of the first two terms on the right-hand side
in the second line of Eq.~(\ref{eq:3.15}) is at least of $O(\beta^8)$ and can be discarded. To evaluate the remaining
three matrix elements we note the identity
\be
\langle e_0^{(1)}\vert \Lambda_3 = -\frac{1}{2}\langle 1\vert \Lambda_4 + 2\,\frac{T_0}{T_1}\,\langle 1\vert\Lambda_2 + O(T_0^4/T_1^3)\ .
\label{eq:3.16}
\ee
This yields
\be
\mu_0^{(4)} \langle 1\vert 1\rangle = 4\,\frac{T_0}{T_1}\,\langle 1\vert\Lambda_2\vert 1\rangle + \langle e_0^{(1)}\vert\Lambda_2\vert e_0^{(1)}\rangle
   + O(\beta^8)
\label{eq:3.17}
\ee
To leading order we further have $K_2 \approx 2(T_0/T_1)K_0$ and thus we can express $\mu_0^{(4)}$ to leading order entirely
in terms of the average value of $\Gamma_0$, Eq.~(\ref{eq:B.2a}). We find
\be
\mu_0^{(4)} = 10(T_0/T_1)^2 \langle\Gamma_0\rangle_{w_{\sigma}} + O(\beta^8)\ .
\label{eq:3.18}
\ee
We will not need the eigenvector to this order.

Combining Eqs.~(\ref{eq:3.11a}) and (\ref{eq:3.18}) we obtain the lowest eigenvalue as
%
\be
\mu_0 = -8\,g_0\,\frac{T_0^3}{T_1\lambda} \left[1 + \frac{T\lambda}{T_0^2} + O\left(\frac{T_0}{T_1}, \frac{T_0\lambda}{T_1^2}, \frac{T\lambda}{T_0T_1}\right)\right] e^{-T_{\text{min}}/T}.
\label{eq:3.19}
\ee
Here we see the mechanism discussed in connection with Eq.~(\ref{eq:3.2}): At asymptotically low temperatures, $T\ll T_0^2/\lambda$, 
the prefactor of the exponential is temperature independent, but in the regime $T_0^2/\lambda \ll T \ll T_0$ the $T_0 T/T_1$ contribution 
from $\mu_0^{(2)}$ dominates.

The right zero eigenvector is
\be
\vert e_0\rangle = \vert 1\rangle + \vert e_0^{(1)}\rangle + \vert e_0^{(2)}\rangle + \vert e_0^{(3)}\rangle + O(\beta^4)
\label{eq:3.20}
\ee
with $\vert e_0^{(1)}\rangle$ from Eq.~(\ref{eq:3.9b}). $\vert e_0^{(2)}\rangle$ and $\vert e_0^{(3)}\rangle$ we have not
determined explicitly, but we know the overlap of $\vert e_0\rangle$ with $\vert\epsilon\rangle$ to lowest order, which is
given by Eq.~(\ref{eq:3.13b}).

We can now construct the leading part of the inverse collision operator. For the matrix elements that determine the
transport coefficients of interest, Eqs.~(\ref{eqs:2.23}), we need to keep only those parts of $\Lambda^{-1}$ that are constructed
from vectors that have an overlap with either $\vert 1\rangle$ or $\vert\epsilon\rangle$. The latter carries a factor of $\alpha^2$
in our power-counting scheme, and the leading scaling behavior of $\Lambda^{-1}$ thus is
\bse
\label{eqs:3.21}
\bea
\Lambda^{-1} &\sim& \frac{1}{\beta^2\alpha^2 + \beta^6}\Bigl[ \vert 1\rangle\langle 1 \vert + \beta\alpha^2 \vert 1\rangle\langle \epsilon\vert
   + \beta \alpha^2 \vert\epsilon\rangle\langle 1\vert 
   \nonumber\\
   && \hskip 60pt + \beta^2\alpha^4\vert \epsilon\rangle\langle \epsilon\vert \Bigr]\,\frac{1}{\langle 1\vert 1\rangle}\ .
\label{eq:3.21a}
\eea
Explicitly, we have
\be
\Lambda^{-1} = \frac{1}{\mu_0} \, \frac{1}{\langle 1\vert 1\rangle} \Bigl[ \vert 1\rangle\langle 1\vert + b \bigl(
     \vert 1\rangle\langle\epsilon\vert - \vert\epsilon\rangle\langle 1\vert\bigr) - b^2 \vert\epsilon\rangle\langle\epsilon\vert \Bigr]\ .
\label{eq:3.21b}
\ee  
\ese
Here $\mu_0$ is the eigenvalue from Eq.~(\ref{eq:3.19}), we have kept only terms that do not vanish upon multiplying from 
either side with $\vert 1\rangle$ or $\vert\epsilon\rangle$, and we have replaced $e_0^{(3)}$ by the effective expressions
from Eqs.~(\ref{eq:3.13d}, \ref{eq:3.13e}). As a result, this expression for the inverse collision operator is adequate only
for solving the integral equations (\ref{eqs:2.21}) to lowest order in our expansion in powers of $T_0$ and $T$.

\subsubsection{Solutions of the integral equations}
\label{subsubsec:III.A.3}

We are now in a position to determine the functions $\varphi_0$ and $\varphi_1$ from Eqs.~(\ref{eqs:2.21}). For $\varphi_0$ we have
\be
\vert\varphi_0\rangle = -\Lambda^{-1}\vert1\rangle = \frac{-1+O(\beta^2)}{\mu_0}\,\vert 1\rangle
\label{eq:3.22}
\ee
For the hydrodynamic contribution to $\varphi_1$ we have
\bse
\label{eqs:3.23}
\be
\vert\varphi_1\rangle^{\text{hyd}} = -\Lambda^{-1}\vert\epsilon\rangle = \frac{-\langle\epsilon^2\rangle_{w_{\sigma}}}{\mu_0}\,\left[b\,\vert 1\rangle
   - b^2\,\vert\epsilon\rangle \right]
\label{eq:3.23a}
\ee
with $b$ from Eq.~(\ref{eq:3.13c}). In addition, there is the kinetic contribution
\be
\vert\varphi_1\rangle^{\text{kin}} = \vert h\rangle
\label{eq:3.23b}
\ee
\ese
with $\vert h\rangle$ the solution of Eq.~(\ref{eq:2.27}). 

$\vert h\rangle$ is hard to determine explicitly, but we can investigate its scaling behavior in order to compare
with the hydrodynamic part. $h$ must be odd in $\epsilon$, and an obvious lowest-order variational {\it ansatz} is
$h_{\sigma}(\epsilon) = h_1\epsilon$, with $h_1$ a spin-independent constant. Eq.~(\ref{eq:2.27}) then yields
(we note again that we do not distinguish between $T_1$ and $\lambda$ for scaling purposes)
\bea
h_1 &=& \frac{-\langle\epsilon\vert\epsilon\rangle}{\langle\epsilon\vert\Lambda_0\vert\epsilon\rangle} \sim \frac{\sqrt{T_0/T_1}}{T_1}\,\frac{\langle\epsilon\vert\epsilon\rangle}{\langle\epsilon\vert\Lambda_3\vert 1\rangle}
\nonumber\\
&\sim& \frac{\sqrt{T_0/T_1}}{T_1}\,\frac{\langle\epsilon^2\rangle_{w_{\sigma}}}{\langle\epsilon\Gamma_3\rangle_{w_{\sigma}}}\ .
\label{eq:3.24}
\eea
where we have used Eq.~(\ref{eq:3.6f}) and $\Gamma_3$ is one of the relaxation rates defined in Eq.~(\ref{eq:2.12}). 
With the help of Eq.~(\ref{eq:B.3}) we have
\bse
\label{eqs:3.25}
\be
\frac{1}{T}\langle\epsilon\vert\varphi_1\rangle^{\text{kin}} \sim \frac{T^4}{T_0^3}\,e^{\Tmin/T}
\label{eq:3.25a}
\ee
This competes with
\be
\frac{1}{T} \langle\epsilon\vert\varphi_1\rangle^{\text{hyd}} \sim \frac{T^4}{T_1 T_0^2}\,e^{\Tmin/T}
\label{eq:3.25b}
\ee
\ese
We see that the kinetic and hydrodynamic contributions have the same temperature scaling, but the latter
is smaller then the former by a factor of $T_0/T_1 = \lambda^2/4\epsilonF^2 \sim T_1^2/\epsilonF^2$. 
Furthermore, an inspection shows that $\langle\epsilon\vert\varphi_1\rangle^{\text{kin}}$ is negative,
and thus gives a positive contribution to the heat conductivity, whereas $\langle\epsilon\vert\varphi_1\rangle^{\text{hyd}}$ 
is positive. This is exactly the same behavior as in the case of intraband (e.g., phonon) scattering, see
Eqs.~(3.39) in Paper I.

\subsection{The transport coefficients}
\label{subsec:III.B}

We now can determine the leading contributions to the transport coefficients. Equations (\ref{eq:2.23a}),
(\ref{eq:3.22}), and (\ref{eq:3.19}) yield, for the electrical conductivity,
\bea
\sigma &=& \frac{n e^2}{m}\,\frac{-1}{\mu_0} 
\nonumber\\
&=&  \frac{n e^2}{m}\,\frac{T_1\lambda}{8 g_0 T_0^3} 
\frac{e^{T_{\text{min}}/T}}{1 + \frac{T\lambda}{T_0^2} + O\left(\frac{T_0}{T_1}, \frac{T_0\lambda}{T_1^2}, \frac{T\lambda}{T_0T_1}\right)}\ .
\nonumber\\
\label{eq:3.26}
\eea
We see that for temperatures $T\ll T_0^2/\lambda$ the prefactor of the exponential is temperature independent
and proportional to $T_1\lambda/T_0^3$, but for $T_0^2/\lambda \ll T \ll T_0$ it is proportional to $T_1/T_0 T$.
The prefactor is exact to the order indicated.

For the thermopower, Eqs.~(\ref{eq:2.23b}), (\ref{eq:3.23a}), and (\ref{eq:3.13c}) yield
\be
-e\,S = \frac{\pi^2}{6}\,\frac{T}{\epsilonF}\ .
\label{eq:3.27}
\ee
This is the same result we obtained for intraband scattering (phonons, antiferromagnons, helimagnons) in Paper I.
The prefactor is again exact.
Note that $\langle h\vert 1\rangle = 0$, so the kinetic part of $\varphi_1$ does not contribute to the thermopower.

For the heat conductivity, we find from Eqs.~(\ref{eq:2.23b}), (\ref{eq:3.23a}), and (\ref{eq:3.13c})
\be
\sigma_h = \frac{n}{mT} \left(\frac{1}{T}\langle h\vert\epsilon\rangle + \frac{b^2 \langle\epsilon^2\rangle_{w_{\sigma}}}{\mu_0}\right)
\label{eq:3.28}
\ee
The result for the heat conductivity is often given as an expression for $\sigma_h/T$, which is proportional to the heat
diffusivity (assuming a specific heat that is linear in $T$) and which dimensionally is an inverse
rate, as is the electrical conductivity. From Eq.~(\ref{eq:3.28}) we find
\bse
\label{eqs:3.29}
\be
\sigma_h/T = \frac{n}{m}\,\frac{1}{g_0} \left(\eta - \frac{\pi^4}{72}\,\frac{T_0}{\lambda}\right)\,\frac{T^2}{T_0^3}\,e^{\Tmin/T}
\label{eq:3.29a}
\ee
where
\be
\eta = g_0\,\frac{T_0^3}{T^5}\,e^{-\Tmin/T} \langle h\vert\epsilon\rangle
\label{eq:3.29b}
\ee
\ese
is a number independent of $T$ and $g_0$. An explicit determination of $\eta$ requires solving the integral
equation (\ref{eq:2.27}). This is the same situation as in the intraband case, see Eq.~(3.39a) in Paper I:
The hydrodynamic contribution to the heat conductivity can be found exactly in closed form, but the 
kinetic part involves a number given as an integral over a scaling function that we have been unable to
determine explicitly.

\section{Summary, and Discussion}
\label{sec:IV}

In summary, we have provided an exact solution of the electron-ferromagnon scattering problem at
low temperatures at the level of the linearized Boltzmann equation or the equivalent conserving
approximation of the Kubo formula, in analogy to the exact solution of the electron-phonon problem in Paper I. While
it is physically obvious that the magnon contributions to the electrical and heat conductivities are 
exponentially large, determining the temperature dependence of the prefactor of the exponential
proved to be a hard problem. The result is a $T$-independent prefactor for the electrical
conductivity, Eq.~(\ref{eq:3.26}), and a $T^3$ behavior for the heat conductivity, Eqs.~(\ref{eqs:3.29}).
The thermopower is linear in $T$, Eq.~(\ref{eq:3.27}). Our method also yields the exact numerical
prefactors. In conclusion, we discuss several aspects of our method and our results.

\subsection{Technical aspects}
\label{subsec:IV.A}

It is worth emphasizing the generality of our method. It relies solely on the existence of a perturbed
zero eigenvalue of the collision operator,\cite{Dorfman_vanBeijeren_Kirkpatrick_2020} which in turn
relies only on the asymptotic conservation of the electron momentum in the limit $T\to 0$. The
low-temperature limit thus provides perturbative control that is not available in classical kinetic
theory. As a result, the transport coefficients can be determined exactly, provided the leading
hydrodynamic contribution to the spectrum of the collision operator (i.e., the one related to the
perturbed zero eigenvalue) dominates the leading kinetic contribution. This is the case for the
electrical conductivity and the thermopower. In the case of the heat conductivity, the hydrodynamic
and kinetic contributions both contribute to the leading term, and an explicit determination of the
kinetic contribution to the numerical prefactor ($\eta$ in Eqs.~(\ref{eqs:3.29}) requires the solution 
of an integral equation that is not amenable to perturbative techniques. These aspects are all
qualitatively the same as in the electron-phonon problem, see Paper I. This illustrates that the
technique is independent of the origin and the nature of excitations that mediate the electron
scattering. In particular, it works equally well for particle-like excitations and for continuum excitations 
that are not characterized by weakly damped poles in the effective potential.

These structural similarities notwithstanding, the ferromagnon problem is harder to solve than
the phonon problem for two reasons: First, the leading temperature dependence of the conductivities is
exponential, and the prefactor is a subleading term. Second, the gap in the effective potential introduces
a new energy scale $T_0$, and it is difficult to distinguish between powers of $T$ and powers
of $T_0$. We solved this problem by means of a double expansion in powers of $T/T_1$ and
$T_0/T_1$, with $T_1$ the magnetic Debye temperature. The first problem is aggravated by the
fact that the leading contributions to the perturbed zero eigenvalue cancel, see Eqs.~(\ref{eqs:3.11}),
which forces one to go to higher order in the double expansion. 

Our exact result for the electrical conductivity differs from the one obtained in
Ref.~\onlinecite{Bharadwaj_Belitz_Kirkpatrick_2014}, which found a $1/T^2$ dependence of the
prefactor of the exponential, rather than the correct $T$-independent result. This discrepancy can
be traced to the common approximation for solving the Boltzmann equation that was used in this reference, which replaces
all relaxation rates by their on-shell values to turn the integral equation in to an algebraic one.\cite{Mahan_2000} 
In the electron-phonon case, this gives the qualitatively correct answer, as was demonstrated in Paper I. 
In the ferromagnon case is does not, since it mistakes powers of $T$ for powers of $T_0$. If one 
replaces the relaxation rates by their energy averages with the appropriate weight function ($w$ in Eq.~(\ref{eq:2.7a})), 
then the algebraic equation gives the qualitatively correct answer, but this statement requires
knowledge of the exact solution. 

In the context of the electron-phonon problem it is sometimes stated that the heat conductivity has the same
temperature dependence as the inverse single-particle relaxation rate, the reason being that
energy relaxation is more isotropic than momentum relaxation and hence not suppressed by
the dominance of backscattering events.\cite{Ziman_1960} As pointed out in Paper I, such statements are misleading, 
and this is particularly obvious in the ferromagnon scattering problem we have discussed: The heat
diffusivity scales as $\sigma_h/T \sim (T^2/T_0^3) e^{\Tmin/T}$, see Eq.~(\ref{eq:3.29a}), 
whereas the inverse single-particle rate scales as $1/\langle\Gamma_0\rangle \sim (1/T_0) e^{\Tmin/T}$, 
see Eq.~(\ref{eq:B.2a}). 

In the context of the heat conductivity it is illustrative to return to the first point in the current subsection.
In order for our method to be controlled, it is crucial that the hydrodynamic eigenvalue of the
collision operator is the smallest one. This notion is indeed consistent with the result for the
heat conductivity. Consider the operator 
\bse
\label{eqs:4.1}
\be
\Lambda_{0\perp} = P_{\perp}\Lambda_0 P_{\perp}
\label{eq:4.1a}
\ee
where 
\be
P_{\perp} = \mathbbm{1} - \frac{\vert 1\rangle\langle 1\vert}{\langle 1\vert 1\rangle}
\label{eq:4.1b}
\ee
\ese
is a projection operator that projects out the subspace spanned by the zero eigenvector of $\Lambda_0$
(see Sec.~III.A in Paper I). $\Lambda_{0\perp}$ has a low-energy representation
\be
\Lambda_{0\perp} \approx \lambda_0\, \frac{\vert \epsilon\rangle\langle \epsilon\vert}{\langle \epsilon\vert \epsilon\rangle}
\label{eq:4.2}
\ee
with $\lambda_0$ the smallest kinetic eigenvalue. This yields
\be
\lambda_0 \approx \frac{\langle\epsilon\vert\Lambda_{0\perp}\vert\epsilon\rangle}{\langle\epsilon\vert\epsilon\rangle} = \frac{1}{h_1} 
\sim \frac{T_0^3}{T^2}\,e^{-\Tmin/T}
\label{eq:4.3}
\ee
This needs to be compared with the hydrodynamic eigenvalue $\mu_0$, which scales as (see Eq.~(\ref{eq:3.19}))
\be
\mu_0 \sim \frac{T_0^3}{T_1^2}\,e^{-\Tmin/T}\ .
\label{eq:4.4}
\ee
We see that for $T\to 0$ we do indeed have $\lambda_0 \gg \mu_0$, which is crucial for our method to work.
We also note that $\lambda_0$ still vanishes as $T\to 0$, just not as fast as $\mu_0$.

\subsection{Observational aspects}
\label{subsec:IV.B}

A semi-quantitative discussion of the observable consequences of electron-ferromagnon scattering has been
given in Ref.~\onlinecite{Bharadwaj_Belitz_Kirkpatrick_2014}, and here we restrict ourselves to a few remarks.

First of all, it is important to remember that there are many contributions to the transport coefficients from
scattering by excitations that are not subject to the interband restriction characteristic of ferromagnons. 
These lead to power-law contributions that dominate at low temperatures and will have to be subtracted
in order to extract the magnon contribution from any transport data.
Second, the asymptotic temperature regime $T\ll T_0$ where our solution is valid is quite low in most materials.
Estimates for $T_0$ in Ref.~\onlinecite{Bharadwaj_Belitz_Kirkpatrick_2014} range from about 30 K in Fe to about
10 mK in Ni$_3$Al. Finally, we mention that we have, strictly speaking, not considered the true asymptotic low-temperature
regime. A more detailed analysis of the temperature scale $\Tmin$ in Eq.~(\ref{eq:2.14}) shows that $\Tmin$ is
spin dependent, and the difference between the two scales in on the order of $\Tmin^+ - \Tmin^- \propto T_0^2/\epsilonF$.
Typical values of the ratio $T_0/\epsilonF$ in metals are on the order of $10^{-7}$,\cite{Bharadwaj_Belitz_Kirkpatrick_2014}
which makes this effect unobservably small.

\appendix

\section{The effective potential, and the structure of the kinetic equations}
\label{app:A}

\subsection{The effective potential}
\label{app:A.1}

In this Appendix we explain the origin of the effective potential given in Eq.~(\ref{eq:2.13b}), and the
structure of the integral equations (\ref{eqs:2.8}).

In a ferromagnet, the split Fermi surface is defined by
\be
\xi_{\sigma}({\bm k}) \equiv \epsilon_{\bm k} - \mu + \sigma\lambda = 0\ ,
\label{eq:A.1}
\ee
where $\epsilon_{\bm k}$ is the single-electron energy, $\lambda$ is the exchange splitting, $\mu$ is
the chemical potential, and $\sigma = \pm$ is the spin index. The effective potential for the electron-electron
interaction mediated by magnon exchange was derived in Ref.~\onlinecite{Bharadwaj_Belitz_Kirkpatrick_2014}
(see also Appendix A in Paper I). It is spin dependent, and proportional to the magnon susceptibility,
\bse
\label{eqs:A.2}
\be
V_{\sigma\sigma'}({\bm k},z) \propto (1 - \delta_{\sigma\sigma'}) \chi_{\sigma'}({\bm k},z)\ ,
\label{eq:A.2a}
\ee
where $z$ is the complex frequency. The susceptibility has the form
\be
\chi_{\pm}({\bm k},z) \propto \frac{1}{\omega_{\bm k} \pm z}
\label{eq:A.2b}
\ee
\ese
with $\omega_{\bm k}$ the magnon resonance frequency from Eq.~(\ref{eq:2.1}). 
The potential $V_{\sigma\sigma'}(u)$ given in Eq.~(\ref{eq:2.13b}) is obtained by
averaging the spectrum $V''$ of $V_{\sigma\sigma'}({\bm k},z)$ over the split Fermi surface,
\bea
V_{\sigma\sigma'}(u) &=& \frac{1}{\NF^{\sigma}\NF^{\sigma'} V^2} \sum_{{\bm k},{\bm p}} \delta(\xi_{\sigma}({\bm k}) - \epsilon)
\delta(\xi_{\sigma'}({\bm p}) - \epsilon - u)\,
\nonumber\\
&&\hskip 80pt \times V''_{\sigma\sigma'}({\bm k}-{\bm p},u)\ .
\label{eq:A.3}
\eea
Performing the integrals yields Eq.~(\ref{eq:2.13b}) with the lower frequency cutoff given by Eq.~(\ref{eq:2.14}).

\subsection{The structure of the kinetic equations}
\label{app:A.2}

\begin{figure}[t]
\vskip -0mm
\includegraphics[width=8.0cm]{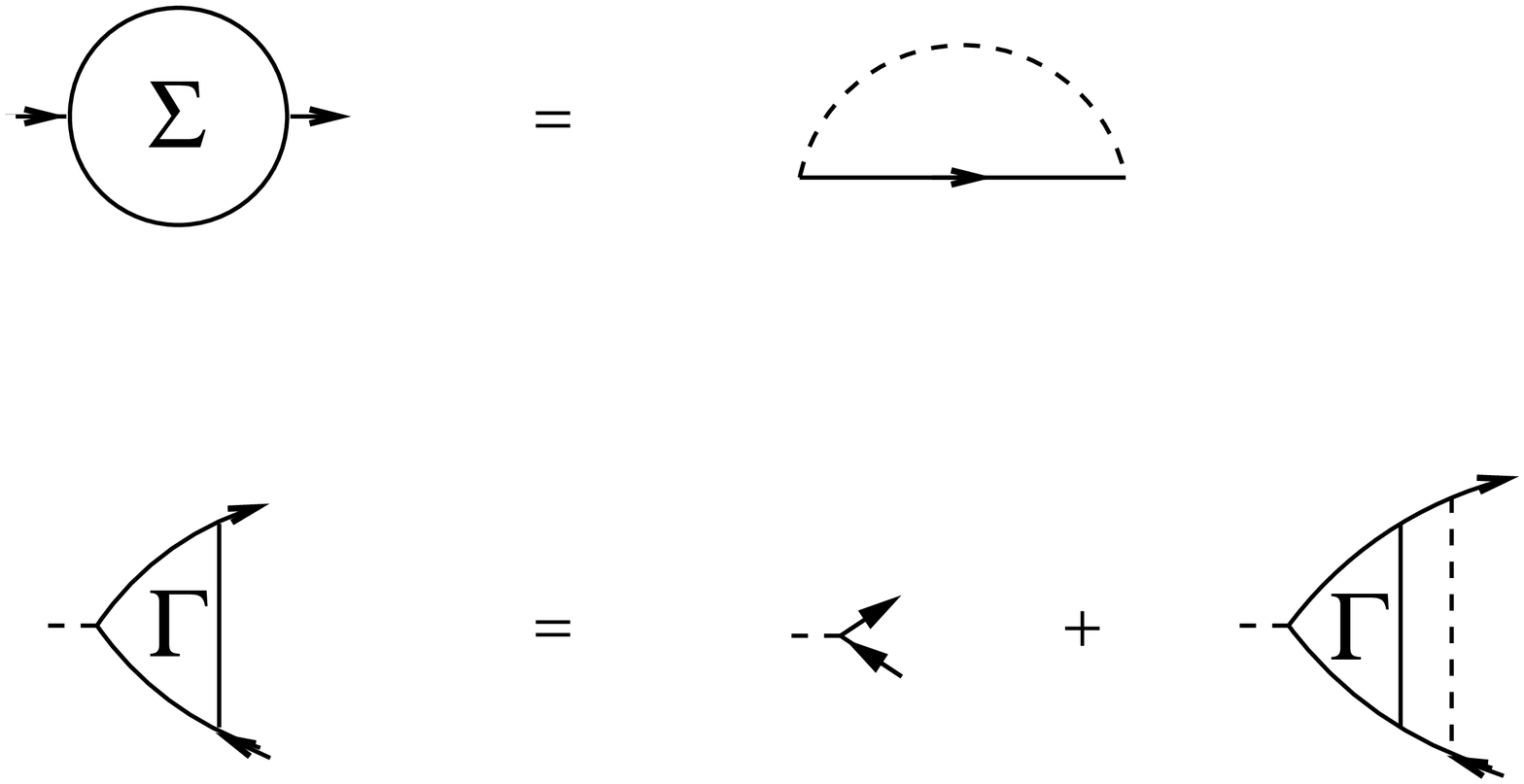}
\caption{Conserving approximation for the self energy $\Sigma$ and the vertex function ${\bm\Gamma}$. Directed solid
   lines denote the Green function $G$, and dashed lines denote the effective potential $V$.}
\label{fig:A.1}
\end{figure}

The electrical conductivity as a function of the imaginary frequency $i\Omega$ is given by the Kubo formula\cite{Kubo_1957, Mahan_2000}
\bse
\label{eqs:A.4}
\be
\sigma_{ij}(i\Omega)=\frac{i}{i\Omega}[\pi_{ij}(i\Omega)-\pi_{ij}(i\Omega=0)]\ ,
\label{eq:A.4a}
\ee
where the tensor
\bea
\pi_{ij}(i\Omega) &=& \frac{-e^2\,T}{m^2}\sum_{i\omega,i\omega'}\frac{1}{V} \sum_{{\bm k},{\bm p}}
k_i\,p_j \sum_{\sigma,\sigma'}
\nonumber\\
&&\hskip -50pt \times \left\langle{\bar\psi}_{\sigma}({\bm
k},i\omega)\,\psi_{\sigma}({\bm k},i\omega+i\Omega)\,{\bar\psi}_{\sigma'}({\bm
p},i\omega')\,\psi_{\sigma'}({\bm p},i\omega'-i\Omega)\right\rangle. \nonumber\\
\label{eq:A.4b}
\eea
\ese
is the current-current susceptibility or polarization function. Here $\psi$ and $\bar\psi$ denote
fermionic fields, $\omega$ and $\Omega$ are fermionic and bosonic Matsubara frequencies,
respectively, and the average is to be taken with respect to an action of electrons that interact
via the effective dynamical potential given in Eqs.~(\ref{eqs:A.2}). The four-fermion correlation
function in Eq.\ (\ref{eq:A.4b}) is conveniently expressed in terms of the single-particle Green function
\be
G_{\sigma}({\bm k},i\omega) = 1/(i\omega - \xi_{\sigma}({\bm k}) - \Sigma_{\sigma}({\bm k},i\omega))
\label{eq:A.5}
\ee
and a vector vertex function ${\bm\Gamma}_{\sigma}$ with components $\Gamma_{\sigma}^i$:
\bea
\pi_{ij}(i\Omega) &=& -ie^2T \sum_{i\omega} \frac{1}{V} \sum_{{\bm p},\sigma} \frac{p^i}{m} G_\sigma({\bm p},i\omega) 
   G_\sigma({\bm p},i\omega-i\Omega) 
   \nonumber\\
   && \hskip 70pt \times \Gamma^j_\sigma({\bm p};i\omega,i\omega-i\Omega)\ .
\label{eq:A.6}
\eea
It is important to calculate the vertex function ${\bm\Gamma}$ and the self energy $\Sigma$ in mutually
consistent approximations.\cite{Kadanoff_Baym_1962}  We use the familiar procedure that consists
of a self-consistent Born approximation for the self energy, and a ladder approximation for the vertex function,
\bea
{\bm\Gamma}_{\sigma}({\bm p};i\omega,i\omega-i\Omega) &=& i\frac{{\bm p}}{m} + \frac{T}{V}\sum_{{\bm k},i\Omega^\prime} \sum_{\sigma'} V_{\sigma\sigma'}({\bm k}-{\bm p},i\Omega')
\nonumber\\
&&   \hskip -60pt \times  G_{\sigma'}({\bm k},i\omega +i\Omega^\prime)\,
                    G_{\sigma'}({\bm k},i\omega - i\Omega + i\Omega^\prime) 
   \nonumber \\
&& \hskip -40pt \times {\bm\Gamma}_{\sigma'}({\bm k};i\omega + i\Omega^\prime,i\omega - i\Omega + i\Omega^\prime)\ .
\label{eq:A.7}
\eea
These approximations are graphically represented in Fig.~\ref{fig:A.1}.
It is convenient to define a scalar vertex function $\gamma$ by 
${\bm\Gamma}({\bm p};i\omega,i\omega') = i({\bm p}/m)\,\gamma({\bm p});i\omega,i\omega')$.
$\gamma$ then obeys
\bea
\gamma_{\sigma}({\bm p};i\omega,i\omega-i\Omega) &=& 1 + \frac{T}{V}\sum_{{\bm k},i\Omega^\prime} \sum_{\sigma'}
  \, V_{\sigma\sigma'}({\bm p} - {\bm k},i\Omega^\prime)
\nonumber\\
&&\hskip -80pt \times  \frac{{\bm p}\cdot{\bm k}}{{\bm p}^2}\,G_{\sigma'}({\bm k},i\omega + i\Omega^\prime)\,
   G_{\sigma'}({\bm k},i\omega - i\Omega + i\Omega^\prime)  
\nonumber\\
&& \hskip -40pt \times    \gamma_{\sigma'}({\bm k};i\omega-i\Omega^\prime,i\omega-i\Omega-i\Omega^\prime)\ .
\label{eq:A.8}
\eea

The polarization and conductivity tensors are diagonal, $\sigma_{ij}(i\Omega) = \delta_{ij}\,\sigma(i\Omega)$, and the sum
over Matsubara frequencies in Eq.\ (\ref{eq:A.6}) can be transformed into an integral along the real axis. In the limit of
low temperature, the imaginary part of the self energy goes to zero, and the real part just renormalizes the Fermi energy. 
The relevant limit is thus the one of a vanishing self energy, and in this limit the leading contributions to the integral come 
from terms where the frequency arguments of the two Green functions lie on different sides of the real axis. In the static limit, 
the Kubo formula for the conductivity $\sigma = \lim_{\Omega\to 0} \Re\sigma(i\Omega\to\Omega + i0)$, thus becomes
\bea
\sigma &=& \frac{e^2}{3\pi m^2}\int_{-\infty}^{\infty} \frac{d\epsilon}{4T} \frac{1}{\cosh^2(\epsilon/2T)} \frac{1}{V} \sum_{{\bm p}} {\bm p}^2
\nonumber\\
&&\times  \sum_{\sigma} \vert{\cal G}_{\sigma}({\bm p},\epsilon+i0)\vert^2 \gamma_{\sigma}({\bm p};\epsilon+i0,\epsilon-i0)\ .\qquad
\label{eq:A.9}
\eea
The pole of the Green function ensures that the dominant contribution to the momentum integral comes from the momenta
that obey $\xi_{\sigma}({\bm p}) = \epsilon$. Furthermore, since $\epsilon$ scales as $T$, for the leading $T$-dependence
we can neglect all $\epsilon$-dependencies that do not occur in the form $\epsilon/T$. Equation (\ref{eq:A.9}) then reduces to
Eq.~(\ref{eq:2.6a}), with the relaxation rate $\varphi_0$ the solution of
\be
\varphi_0^{\sigma}(\epsilon)\,\Gamma_0^{\sigma}(\epsilon) = 1 - \int du \sum_{\sigma'} K^{\sigma\sigma'}(\epsilon,u)\,\varphi_0^{\sigma'}(u)
\label{eq:A.10}
\ee
with $K$ as defined in Sec.~\ref{subsec:II.B} and $\Gamma_0$ from Eq.~(\ref{eq:2.12}). Here we have used the fact that the 
Green functions in Eq.~(\ref{eq:A.8}) pin the wave vectors ${\bm k}$ and ${\bm p}$ to energy shells at distances $\epsilon\approx 0$ 
and $\epsilon + u \approx u$, respectively, from the Fermi surface, see Eq.~(\ref{eq:A.3}). As a result, the factor ${\bm p}\cdot{\bm k}/{\bm p}^2$ in the 
integrand of Eq.~(\ref{eq:A.8}) effectively becomes
\be
\frac{{\bm p}\cdot{\bm k}}{{\bm p}^2} = 1 - \frac{{\bm p}\cdot({\bm p}-{\bm k})}{{\bm p}^2} \to 1 - \frac{1}{2{\bm p}^2}({\bm p}-{\bm k})^2 + \frac{mu}{{\bm p}^2}\ .
\label{eq:A.11}
\ee
The second term on the far right-hand side is what is often called the ``backscattering factor'' in transport theory. It gives rise to the
kernel $K_2$. The last term gives rise to the kernel $K_3$. $K_1$, the second contribution to $K_3$, and $K_4$ arise from the
spin dependence of the density of states. 

Equation (\ref{eq:A.10}) has the structure of the kinetic equation discussed in Ref.~\onlinecite{Bharadwaj_Belitz_Kirkpatrick_2014}.
It can obviously be rewritten in the form of Eq.~(\ref{eq:2.8a}). Note that Ref.~\onlinecite{Bharadwaj_Belitz_Kirkpatrick_2014}
ignored $K_4$ and the first contribution to $K_3$, an approximation that cannot be justfied {\em a priori}.\cite{K34_footnote}

The equation for the relaxation function $\varphi_1$ is obtained by the same procedure, with the number current ${\bm p}/m$
in Eqs.~(\ref{eq:A.6}) and (\ref{eq:A.7}) replaced by the heat current ${\bm p}\,\xi_{\sigma}({\bm p})/m$. This leads to Eq.~(\ref{eq:A.10})
with the $1$ on the right-hand side replaced by $\epsilon$, and hence to Eq.~(\ref{eq:2.8b}).

\section{Relaxation rates}
\label{app:B}

The lowest eigenvalue of the collision operator calculated in Sec.~\ref{sec:III} depends on the relaxation rates
$\Gamma_n$, Eq.~(\ref{eq:2.12}), averaged according to Eq.~(\ref{eq:2.20}). In order to calculate the averages
it is advantageous to do the $\epsilon$ integration first, making use of the integral
\be
\frac{1}{T} \int d\epsilon\,w(\epsilon)\,f_0(x \mp \epsilon/T) = \frac{-1}{e^x-1} + \frac{x\,e^x}{(e^x-1)^2}
\label{eq:B.1}
\ee
which is independent of the sign in the argument of the Fermi function $f_0$. We obtain
%
\bse
\label{eqs:B.2}
\bea
\langle\Gamma_0\rangle_{w_{\sigma}} &=& {\tilde g}_0 (\Tmin + T) e^{-\Tmin/T}\ ,
\label{eq:B.2a}\\
\langle\Gamma_2\rangle_{w_{\sigma}} &=& \frac{{\tilde g}_0}{1-4T_0/T_1} \left[-2\left(\frac{\Tmin^2}{T_1} + 2\,\frac{\Tmin T}{T_1} + 2\,\frac{T^2}{T_1}\right) \right.
\nonumber\\
&& \hskip 35pt \left.     + \frac{4T_0}{T_1}\left(\Tmin + T\right)\right]\,e^{-\Tmin/T}\ ,
\label{eq:B.2b}\\
 \langle\Gamma_4\rangle_{w_{\sigma}} &=& \frac{2 {\tilde g}_0\,T_0/\lambda}{1-4T_0/T_1} \left(\frac{\Tmin^2}{T_1} + 2\,\frac{\Tmin T}{T_1} + 2\,\frac{T^2}{T_1}
 \right)
 \nonumber\\
 && \hskip 80pt \times\,e^{-\Tmin/T}
\label{eq:B.2c}
\eea
\ese    
Here ${\tilde g}_0 = g_0 8\kF^+ \kF^-/\kF^2\nu$ with $\nu$ from Eq.~(\ref{eq:2.17c}). The averages of $\Gamma_1$ and
$\Gamma_3$ vanish by symmetry. Also useful is the average
\be
\langle\epsilon\,\Gamma_3\rangle_{w_{\sigma}} = {\tilde g}_0 \left(\frac{4}{T_1} - \frac{1}{\lambda}\right) T_0^3 \sqrt{\frac{T_0}{T_1}} \left[1 + O(T_0/T_1)\right] e^{-\Tmin/T}\ .
\label{eq:B.3}
\ee

\section{Properties of the zero eigenvector}
\label{app:C}

In this appendix we determine the overlap of the zero eigenvector $\vert e_0\rangle$ with $\vert\epsilon\rangle$ to lowest
order in $\beta \sim \sqrt{T_0/T_1}$. In order to prove Eqs.~(\ref{eq:3.11c}) and (\ref{eq:3.13b}) it is advantageous to rewrite
the eigenproblem (\ref{eq:3.5}) by expanding the collision operator strictly in powers of $\beta \sim \sqrt{T_0/T_1}$. We
define a collision operator
\bse
\label{eqs:C.1}
\be
\tilde\Lambda_0^{\sigma}(\epsilon) =  \int du \sum_{\sigma'} \left[ {\tilde K}_0^{\sigma\sigma'}(\epsilon,u)\,R_{\epsilon\to u}^{\sigma\to\sigma'}   
   - {\tilde K}_0^{\sigma\sigma'}(\epsilon,u) \right]\ .
\label{eq:C.1a}
\ee
in terms of a kernel 
\be
{\tilde K}_0^{\sigma\sigma'}(\epsilon,u) = (\NF/\NF^{\sigma'}) K_0^{\sigma\sigma'}(\epsilon,u)\ ,
\label{eq:C.1b}
\ee
\ese
with $K_0$ from Eq.~(\ref{eq:2.13a}). Similarly, we define
\bse
\label{eqs:C.2}
\be
\tilde\Lambda_1^{\sigma}(\epsilon) = -2\left(\frac{T_0}{T_1}\right)^{1/2}\sigma\,\tilde\Lambda_0^{\sigma}(\epsilon)
\label{eq:C.2a}
\ee
and
\bea
\tilde\Lambda_2^{\sigma}(\epsilon) &=&  \int du \sum_{\sigma'} {\tilde K}_2^{\sigma\sigma'}(\epsilon,u)\,R_{\epsilon\to u}^{\sigma\to\sigma'}  
   + 2\,\frac{T_0}{T_1}\, \tilde\Lambda_0^{\sigma}(\epsilon)
   \nonumber\\
\label{eq:C.2b}\\
\tilde\Lambda_3^{\sigma}(\epsilon) &=&  \int du \sum_{\sigma'}{\tilde K}_3^{\sigma\sigma'}(\epsilon,u)\,R_{\epsilon\to u}^{\sigma\to\sigma'}  
\nonumber\\
&&\hskip 50pt    - 4 \left(\frac{T_0}{T_1}\right)^{3/2} \sigma\,\tilde\Lambda_0^{\sigma}(\epsilon)
\label{eq:C.2c}
\eea   
\ese
where
\bse
\label{eqs:C.3}
\bea
{\tilde K}_2^{\sigma\sigma'}(\epsilon,u) &=& -2\left(\frac{\vert u-\epsilon\vert}{T_1} - \frac{T_0}{T_1}\right) {\tilde K}_0^{\sigma\sigma'}(\epsilon,u)
\label{eq:C.3a}\\
{\tilde K}_3^{\sigma\sigma'}(\epsilon,u) &=& \left(\frac{T_0}{T_1}\right)^{1/2} \left(\frac{u-\epsilon}{\lambda} + 4\,\frac{\vert u-\epsilon\vert}{T_1}\,\sigma
   - 4\,\frac{T_0}{T_1}\,\sigma\right)
\nonumber\\
&& \hskip 80pt \times {\tilde K}_0^{\sigma\sigma'}(\epsilon,u)
\label{eq:C.3b}
\eea
\ese
The right eigenproblem (\ref{eq:3.5}) can then be rewritten as
\bse
\label{eqs:C.4}
\be
\tilde\Lambda^{\sigma}(\epsilon)\,{\tilde e}_0^{\sigma}(\epsilon) = \mu_0\,{\tilde e}_0^{\sigma}(\epsilon)\,\NF/\NF^{\sigma}\ .
\label{eq:C.4a}
\ee
where
\be
\tilde\Lambda = \tilde\Lambda_0 + \tilde\Lambda_1 + \tilde\Lambda_2 + \tilde\Lambda_3 + O(\beta^4)
\label{eq:C.4b}
\ee
and
\be
{\tilde e}_0^{\sigma}(\epsilon) = (\NF^{\sigma}/\NF)\,e^{\sigma}(\epsilon)\ .
\label{eq:C.4c}
\ee
\ese
We further define a weight function 
\bse
\label{eqs:C.4'}
\be
{\tilde w}_{\sigma}(\epsilon) = \frac{1}{2}\,w(\epsilon)
\label{eq:C.4'a}
\ee
with $w(\epsilon)$ from Eq.~(\ref{eq:2.7a}) and an associated scalar product
\be
(\psi\vert\varphi) = \int d\epsilon \sum_{\sigma} {\tilde w}_{\sigma}(\epsilon)\,\psi^{\sigma}(\epsilon)\,\varphi^{\sigma}(\epsilon)
\label{eq:C.4'b}
\ee
\ese
in analogy to Eq.~(\ref{eq:2.19}).

The advantage of this formulation is that various functions associated with $\tilde\Lambda_0$ have simple symmetry properties.
For instance, the relaxation rates
\bse
\label{eqs:C.5}
\be
\tilde\Gamma_n^{\sigma}(\epsilon) = \int du \sum_{\sigma'} {\tilde K}_n^{\sigma\sigma'}(\epsilon,u)\qquad (n=0,2)
\label{eq:C.5}
\ee
defined in analogy to Eq.~(\ref{eq:2.12}) obey
\be
\tilde\Gamma_n^{-}(-\epsilon) = \tilde\Gamma_n^{+}(\epsilon) \qquad (n=0,2)
\label{eq:C.5b}
\ee
\ese
Similarly, the function ${\tilde h}$ defined in analogy to Eq.~(\ref{eq:2.27}),
\bse
\label{eqs:C.6}
\be
\tilde\Lambda_0\vert{\tilde h}\rangle = - \vert\epsilon\rangle
\label{eq:C.6a}
\ee
obeys
\be
{\tilde h}^{-}(-\epsilon) = - {\tilde h}^{+}(\epsilon)
\label{eq:C.6b}
\ee
\ese
A disadvantage is the resulting adjoint properties: $\tilde\Lambda_0$ and $\tilde\Lambda_2$ are self-adjoint with respect
to the scalar product defined in Eq.~(\ref{eq:C.4'b}), but $\tilde\Lambda_1$ and $\tilde\Lambda_3$ are neither self-adjoint
nor skew-adjoint. For our current purposes, this is irrelevant, but the symmetry properties expressed in Eqs.~(\ref{eq:C.5b}) and
(\ref{eq:C.6b}) are crucial. 

We now solve the eigenproblem order by order in $\beta$ as in Sec.~\ref{subsubsec:III.A.2}. The eigenvalues are the same,
as they must be. For the eigenvectors we obtain
\be
\vert {\tilde e}_0^{(1)}\rangle = 0\ ,
\label{eq:C.7}
\ee
This is consistent with Eq.~(\ref{eq:3.9b}), as can be seen by using Eq.~(\ref{eq:C.4c}). 
The equation for $\vert {\tilde e}_0^{(2)}\rangle$ reads
\bse
\label{eqs:C.8}
\be
\tilde\Lambda_0\vert{\tilde e}_0^{(2)}\rangle = -\vert\tilde\Gamma_2\rangle + \langle\tilde\Gamma_2\rangle_{\tilde w}\, \vert 1\rangle\ .
\label{eq:C.8a}
\ee
By multiplying from the left with $\langle{\tilde h}\vert$ and using Eqs.~(\ref{eq:C.5b}) and (\ref{eq:C.6b}) we find
\be
\langle\epsilon\vert{\tilde e}_0^{(2)}\rangle = 0\ .
\label{eq:C.8b}
\ee
\ese
Using Eq.~(\ref{eq:C.4c}) yields Eq.~(\ref{eq:3.11c}).

At the next order, analogous arguments yield
\be
\langle\epsilon\vert{\tilde e}^{(3)}\rangle = \frac{-1}{\lambda} \left(\frac{T_0}{T_1}\right)^{1/2} \langle\epsilon\vert\epsilon\rangle\ .
\label{eq:C.9}
\ee
Here we have used the fact that $\mu_0^{(2)}$ is exponentially small and can be neglected. 
The lowest component of $\vert e_0\rangle$ that has a nonzero overlap with $\vert\epsilon\rangle$ is thus 
$\vert e_0^{(3)}\rangle$, and to leading order $\langle\epsilon\vert e_0^{(3)}\rangle = \langle\epsilon\vert{\tilde e}_0^{(3)}\rangle$.
We thus have
\bse
\label{eqs:C.10}
\bea
\langle\epsilon\vert e_0\rangle &=& -\left(\frac{T_0}{T_1}\right)^{1/2} \frac{1}{\lambda}\,\langle\epsilon\vert\epsilon\rangle\ ,
\label{eq:C.10a}\\
\langle e_0\vert\epsilon\rangle &=& \left(\frac{T_0}{T_1}\right)^{1/2} \frac{1}{\lambda}\,\langle\epsilon\vert\epsilon\rangle\ .
\label{eq:C.10b}
\eea
\ese
Note that the left and right eigenvectors are equal and opposite since $\Lambda_3$ (as opposed to $\tilde\Lambda_3$) is skew-adjoint.
This is the result we have used in Eqs.~(\ref{eqs:3.13}).


\begin{thebibliography}{19}
\bibitem[{mag()}]{magnon_dispersion_footnote}
\bibinfo{note}{We consider a purely quadratic magnon dispersion relation for
  simplicity. Higher powers of ${\bm k}^2$ in $\omega_{\bm k}$ will lead to
  additional kernels that contribute to the collision operator in
  Sec.~\ref{sec:II}. These kernels are of higher order in our power-counting
  scheme and can easily be taken into account, if desirable, within the
  framework of our solution scheme.}
\bibitem[{LR_()}]{LR_footnote}
\bibinfo{note}{In order to avoid cumbersome notation we always denote right
  eigenvectors by $\vert e_n\rangle$ and left eigenvectors by $\langle
  e_n\vert$. It is important to remember that these two vectors represent
  different functions. In particular, $\langle f\vert e_n\rangle \neq \langle
  e_n\vert f\rangle$ for a given function $f$. A more precise, if more
  complicated, notation would be to denote the right eigenfunctions by $e_n^R$,
  and the left ones by $e_n^L$, in which case $\langle f\vert e_n^R\rangle =
  \langle e_n^R\vert f\rangle \neq \langle e_n^L\vert f\rangle$.}
\end{thebibliography}

\end{document}